# DeepALM: Holistic Optical Network Monitoring based on Machine Learning

Joo Yeon Cho[1], Jose-Juan Pedreno-Manresa[1], Sai Patri[1], Khouloud Abdelli[1,2], Carsten Tropschug[3], Jim Zou[1], Piotr Rydlichowski[4]

[1]ADVA Optical Networking, Fraunhoferstrasse 9a, Martinsried, Germany, 82152
[2]Christian-Albrechts-Universität zu Kiel, Kaiserstr. 2, Kiel, Germany, 24143
[3]ADVA Optical Networking SE, Märzenquelle 1-3, Meiningen, Germany, 98617
[4]PSNC, Wieniawskiego 17/19, 61-704, Poznań, Poland
*jcho@adva.com*

**Abstract:** We demonstrate a machine learning-based optical network monitoring system which can integrate fiber monitoring, predictive maintenance of optical hardware, and security information management in a single solution. © 2022 The Author(s)

## 1. Overview

Optical network is the essential backbone for the transportation of aggregated internet, mobile backhaul, and core network. A single fiber link connects thousands of customers and enterprises, carrying a mixture of personal, business, and public data. The impact of network outage can be enormous and must be responded to immediately.

A remote fiber monitoring system can automatically diagnose and locate the fault location quickly. This eliminates the time needed to investigate the cause and determine a search area. The optical networks are monitored in an in-band channel, and it does not require additional monitoring layer. Upon finding the fault location, appropriate action is taken to remedy the fault and restore service as quickly as possible.

Machine learning (ML) has recently emerged as a powerful tool to enhance the fiber monitoring and the predictive maintenance for optical devices and thereby, improve network reliability and operational efficiency, and reduce unplanned downtime and management costs.

In this demo, we present a unified approach for the ML-based optical monitoring that can accommodate multiple monitoring tasks such as fiber monitoring and predictive maintenance in a physical layer, and security information management in a management layer. Although the goal of each task is different, their approaches for data processing, anomaly detection and faults identification are similar, so that users can develop various ML models using individual task-dependent datasets by a uniform set of ML algorithms built-in the software.

The benefits of this approach are many; users can get a holistic view on the optical network in operation and identify and localize the incidents such as hardware fault and fiber cut immediately. Users can also evaluate and report a risk of hardware failure and predict upcoming network maintenance. Instead of running multiple software, they can monitor in a central place in the frontend that provides high-level personalized details about the monitored environment. Furthermore, users can maintain the ML algorithms in a single place and train the models in a uniform way.

We implement an ML-based fiber monitoring, optical sensing and predictive maintenance tool operating over ADVA ALM [2], and a security information management and event triggering tool using ADVA FSP3000 [3] as shown in the figure. DeepALM will be designed using an open source-based monitoring software and standard interfaces.

## 2. Innovation

It is challenging to build a flexible and intelligent monitoring system with ML enabled for a large scale of optical network. Many details must be addressed to develop various ML models using different datasets and implement multiple interfaces to integrate heterogeneous devices, coping with various scenarios of the incidents and detecting anomalies in data communication, management, passive devices, and optical sensors.

DeepALM aims at supporting real-time decision making by combining technology including optical sensing, connectivity, automation, machine learning and real-time processing. The ML-enabled monitoring functions enable the precocious diagnosis of fiber faults, hardware failures and cyber/physical attacks. They also allow the diagnosis and the prediction of upcoming maintenance and the implementation of new processes to prevent network disruption, creating a new pipeline of precocious diagnosis followed by preventive actions.

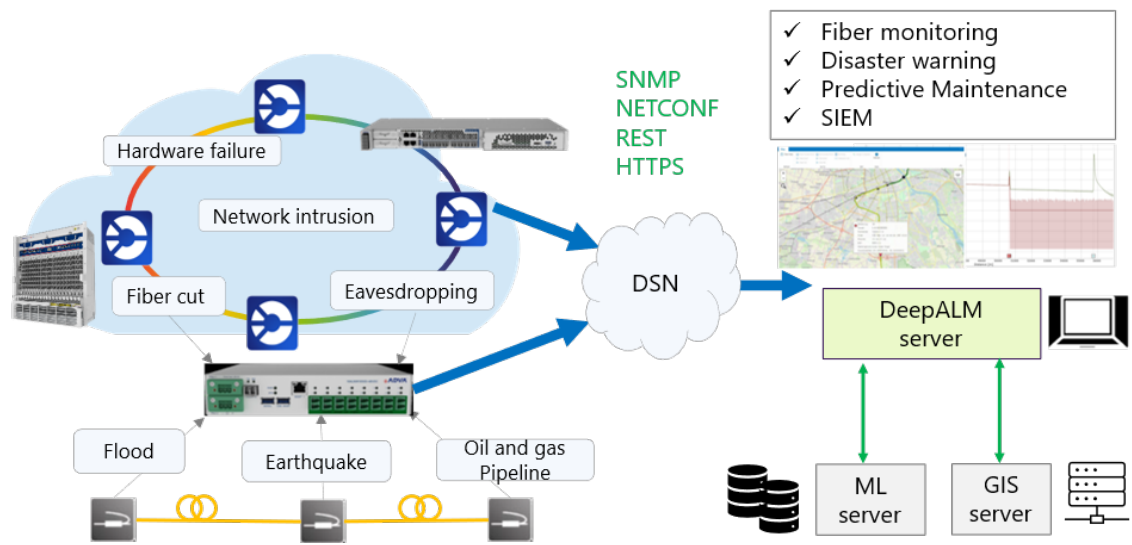

3. **OFC Relevance**

OFC covers aspects of network orchestration and intelligence, hardware, and physical layer transmission. Our demo proposal is well matched with the purpose of the conference, in particular, in the field of the application of AI and ML to optical networking, including autonomous network management and control.

4. **The objectives and the configuration of the demo**

4.1 Fiber monitoring

Fiber monitoring is composed of five main stages: (1) optical fiber network monitoring and data collection, (2) data processing, (3) fiber anomaly detection, (4) fiber fault diagnosis and localization, (5) mitigation and recovery from fiber failures. The optical fibers deployed in the network infrastructure are periodically monitored using the optical time-domain reflectometer (OTDR), a technique based on Rayleigh backscattering, widely applied for fiber characteristics' measurements and for fiber fault detection and localization. In addition, by deploying sensors based on the passive optical fiber (e.g., fiber Bragg grating, bending apparatus), such an OTDR can detect water, gas, vibration, intrusion on premises and other physical alteration. By processing and correlating the detected data, such a system may help proactively alert security breach, nature disaster, and more.

4.2 Predictive maintenance

Optical networks rely on fully functional hardware components that run under optimal conditions. To reduce the risk of unplanned network interruption and service outage, it is important to correctly estimate the degradation of hardware network components using analyzing tools and techniques, by which the maintenance cost and resource allocation are determined.

ML-based prediction is an emerging method to improve the accuracy of estimation of maintenance work for large networks. Since the hardware failures or maintenance events do not occur frequently, it takes time until meaningful training data are collected. Hence, the accelerated aging test results (e.g., a life cycle under the extreme temperature or the over-powered condition) are usually used for training a model.

4.3 Security information and event management (SIEM)

SIEM is a solution that provides monitoring, detection, and alerting of security events or incidents. In this demo, the log data are periodically aggregated throughout equipment in an optical network and analyzed with built-in machine learning algorithms. If incidents and events are identified by an anomaly detection algorithm, the real-time alerts are displayed in dashboards, and the reports are delivered to several critical business and management units.

## 5. How the demonstration will be physically set up

The demo will be composed of two sections: one is a simple optical network using an ALM, FSP3000, optical sensors and fibers, and the other is a monitoring server including an ML server, a GIS server and a dashboard. DeepALM will display the collected data and relevant information in web-based dashboards. It can display metrics, problems, infrastructure, and geometric maps on your dashboards.

## 6. How the demo will be presented to the attendees,

DeepALM will show the capability of ML-based anomaly detection and analysis of optical network data in real-time, enabling users to get holistic insights on the network in operation. Detected problems and incidents can be alerted and tagged for further investigation.

## 7. How attendees might be able to interact with the demonstration (this last feature is desirable to make the session more engaging).

Attendees may have a chance to trigger an incident manually by manipulating an optical sensor or a fiber in the demo. Such incidents will be detected by ALM and analyzed by ML. Then, the suspected root cause and the location of incidents will be identified and displayed in the DeepALM server.

## 8. Acknowledgements

This work has been performed in the framework of the CELTIC-NEXT project AI-NET-PROTECT (Project ID C2019/3-4), and it is partly funded by the German Federal Ministry of Education and Research (FKZ16KIS1279K).